\documentclass{JHEP3}
\usepackage{graphicx}

\catcode`@=11
\newcount\countdp \newcount\countwd \newcount\countht 
\@ifundefined{pdfoutput}{
\def\rgboo#1{}
\def\postscript#1{\special{" #1}}		
\postscript{
        /bd {bind def} bind def
        /fsd {findfont exch scalefont def} bd
        /sms {setfont moveto show} bd
        /ms {moveto show} bd
        /pdfmark where          
        {pop} {userdict /pdfmark /cleartomark load put} ifelse
        [ /PageMode /UseOutlines                
        /DOCVIEW pdfmark}
\def\hhref#1#2{%
        \hskip-.25em\setbox0=\hbox{#1}%
                \countdp=\dp0 \countwd=\wd0 \countht=\ht0%
                \divide\countdp by65536 \divide\countwd by65536%
                        \divide\countht by65536%
                \advance\countdp by1 \advance\countwd by1%
                        \advance\countht by1%
                \def\linkdp{\the\countdp} \def\linkwd{\the\countwd}%
                        \def\linkht{\the\countht}%
        \postscript{
                [ /Rect [ -1.5 -\linkdp.0 0\linkwd.0 0\linkht.5 ] 
                /Border [ 0 0 0 ]
                /Action << /Subtype /URI /URI (#1) >>
                /Subtype /Link
                /ANN pdfmark}{\rgb{.1 .1 1}{#2}}}
}{
\def\rgboo#1{\pdfliteral{#1 rg #1 RG}}
\def\hhref#1#2{%
        \noindent\pdfstartlink user
                {/Subtype /Link
                /Border [ 0 0 0 ]
                /A << /S /URI /URI (#1) >>}{\rgb{.1 .1 1}{#2}}%
        \pdfendlink}
}
\def\rgbo#1#2{\rgboo{#1}#2\rgboo{0 0 0}}
\def\rgb#1#2{\mark{#1}\rgbo{#1}{#2}\mark{0 0 0}}

\def\@spires#1{\hhref{http://www-spires.slac.stanford.edu/spires/find/hep/www?j=#1}} 
\begingroup
\catcode`\ =\active\catcode`,=\active\global
\def\@@keywords#1{\gdef\@keywords{\noindent{\scshape\keywordsname}
		\bgroup\def, {+}\def {_}
		\hhref{http://jhep.sissa.it/stdsearch}%
						{\let,\@comma\let \ #1}.
		\egroup}\egroup\global\@keywordstrue}%
\endgroup
\catcode`@=12
\catcode`\%=12
\renewcommand\jhep[3]  {\hhref{http://jhep.sissa.it/stdsearch?paper=#1
		{{\it J. High Energy Phys.\ }{\bf #1} (#2) #3}}
\catcode`\%=14
\renewcommand{\hepth}[1]{\hhref{http://xxx.lanl.gov/abs/hep-th/#1}{\tt hep-th/#1}}
\renewcommand{\hepph}[1]{\hhref{http://xxx.lanl.gov/abs/hep-ph/#1}{\tt hep-ph/#1}}
\renewcommand{\grqc}[1]{\hhref{http://xxx.lanl.gov/abs/gr-qc/#1}{\tt gr-qc/#1}}
\renewcommand{\astroph}[1]{\hhref{http://xxx.lanl.gov/abs/astro-ph/#1}{\tt astro-ph/#1}}
\renewcommand\email[1]{{\tt\hhref{mailto:#1}{#1}}}

\renewcommand{\theequation}{\thesection.\arabic{equation}}
\newcommand{\be}{\begin{equation}}
\newcommand{\ee}{\end{equation}}
\newcommand{\beqy}{\begin{eqnarray}}
\newcommand{\eeqy}{\end{eqnarray}}
\newcommand{\p}{\partial}
\newcommand{\hp}{\widehat{\p}}
\newcommand{\ov}{\overline}
\newcommand{\da}{^{\dagger}}
\newcommand{\w}{\wedge}
\newcommand{\st}{\stackrel}
\newcommand{\mb}{\mbox}
\newcommand{\mx}{\mbox}
\newcommand{\mt}{\mathtt}
\newcommand{\dt}{\mathtt{d}}
\newcommand{\al}{\alpha}
\newcommand{\bb}{\beta}
\newcommand{\ga}{\gamma}
\newcommand{\Ga}{\Gamma}
\newcommand{\te}{\theta}
\newcommand{\Te}{\Theta}
\newcommand{\de}{\delta}
\newcommand{\De}{\Delta}
\newcommand{\et}{\tilde{e}}
\newcommand{\ze}{\zeta}
\newcommand{\s}{\sigma}
\newcommand{\e}{\epsilon}
\newcommand{\om}{\omega}
\newcommand{\Om}{\Omega}
\newcommand{\la}{\lambda}
\newcommand{\La}{\Lambda}
\newcommand{\n}{\nabla}
\newcommand{\hn}{\widehat{\nabla}}
\newcommand{\hph}{\widehat{\phi}}
\newcommand{\ah}{\widehat{a}}
\newcommand{\bh}{\widehat{b}}
\newcommand{\ch}{\widehat{c}}
\newcommand{\ddh}{\widehat{d}}
\newcommand{\eh}{\widehat{e}}
\newcommand{\gh}{\widehat{g}}
\newcommand{\ph}{\widehat{p}}
\newcommand{\qh}{\widehat{q}}
\newcommand{\mh}{\widehat{m}}
\newcommand{\nh}{\widehat{n}}
\newcommand{\Dh}{\widehat{D}}
\newcommand{\stu}{\st{\textvisiblespace}}
\newcommand{\au}{\stu{a}}
\newcommand{\bu}{\stu{b}}
\newcommand{\cu}{\stu{c}}
\newcommand{\du}{\stu{d}}
\newcommand{\eu}{\stu{e}}
\newcommand{\mmu}{\stu{m}}
\newcommand{\nnu}{\stu{n}}
\newcommand{\pu}{\stu{p}}
\newcommand{\Du}{\stu{D}}
\newcommand{\sto}{\st{\circ}}
\newcommand{\as}{\st{\circ}{a}}
\newcommand{\bs}{\st{\circ}{b}}
\newcommand{\cs}{\st{\circ}{c}}
\newcommand{\ds}{\st{\circ}{d}}
\newcommand{\es}{\st{\circ}{e}}
\newcommand{\ms}{\st{\circ}{m}}
\newcommand{\ns}{\st{\circ}{n}}
\newcommand{\ps}{\st{\circ}{p}}
\newcommand{\Ds}{\st{\circ}{D}}
\newcommand{\sts}{\st{s}}
\newcommand{\sth}{\st{\heartsuit}}
\newcommand{\stp}{\st{\perp}}
\newcommand{\std}{\st{\diamondsuit}}
\newcommand{\ad}{\dot{a}}
\newcommand{\bd}{\st{s}{b}}
\newcommand{\cd}{\st{s}{c}}
\newcommand{\gd}{\st{s}{g}}
\newcommand{\dd}{\st{s}{d}}
\newcommand{\Dd}{\st{s}{D}}
\newcommand{\ed}{\st{s}{e}}
\newcommand{\fd}{\st{s}{f}}
\newcommand{\zd}{\st{s}{\xi}}
\newcommand{\md}{\st{s}{m}}
\newcommand{\nd}{\st{s}{n}}
\newcommand{\stc}{\st{c}}
\newcommand{\az}{\st{c}{a}}
\newcommand{\bz}{\st{c}{b}}
\newcommand{\cz}{\st{c}{c}}
\newcommand{\dz}{\st{c}{d}}
\newcommand{\Dz}{\st{c}{D}}
\newcommand{\ez}{\st{c}{e}}
\newcommand{\fz}{\st{c}{f}}
\newcommand{\nz}{\st{c}{n}}
\newcommand{\mz}{\st{c}{m}}
\newcommand{\tb}{\overline{\theta}}
\newcommand{\ti}{\widetilde}

\newcommand{\hb}{\bar{h}}
\newcommand{\sqw}{\sqrt{w\over 2}\ }

\newcommand{\2}{\frac{1}{2}}
\newcommand{\3}{\frac{1}{3}}
\newcommand{\4}{\frac{1}{4}}
\newcommand{\8}{\frac{1}{8}}
\newcommand{\6}{\frac{1}{16}}

\newcommand{\ra}{\rightarrow}
\newcommand{\Ra}{\Rightarrow}
\newcommand{\im}{\Longleftrightarrow}
\newcommand{\hs}{\hspace{5mm}}
\newcommand{\x}{\star}
\newcommand{\Delt}{\p^{\star}}
\newcommand{\vs}{\vspace{5mm}\\}
\newcommand{\ie}{{\it i.e.}}

\title{Bouncing Universes in  String-inspired Gravity}

\author{Tirthabir Biswas%
\thanks{Supported by NSERC Grant No. 204540.}\\
McGill University, 3600 University Ave., Montr\'eal, Qu\'ebec, Canada\\
\email{biswas@hep.physics.mcgill.ca}}
\author{Anupam Mazumdar\\
NORDITA, Blegdamsvej-17, DK-2100, Copenhagen, Denmark\\
\email{anupamm@nordita.dk}}
\author{Warren Siegel%
\thanks{Supported in part by NSF Grant PHY-0354776.}\\
C.N. Yang Institute for Theoretical Physics, State University of New 
York,\\
Stony Brook, NY 11794-3840, USA\\
\email{siegel@insti.physics.sunysb.edu}}

\abstract{We consider the effects on cosmology of higher-derivative
modifications of (effective) gravity that make it asymptotically free
without introducing ghosts.  The weakening of gravity at short
distances allows pressure to prevent the singularity, producing a
solution with contraction preceding expansion.}

\preprint{hep-th/0508194\\ YITP-SB-05-23\\NORDITA-2005-53}
\begin{document}

\section{Introduction and Summary}

Cosmology acts as a test bed for theories beyond the Standard
Model, including theories for quantum gravity and string theory. The
current observational data strongly favors primordial inflation,
close to the grand unified scale~\cite{WMAP}.  Inflation is a
dynamical solution to the flatness, horizon and homogeneity
problems~\cite{Linde}. In a generic model, such as one driven by a
scalar field, it also stretches metric fluctuations outside the Hubble
radius with a scale invariant power spectrum~\cite{Brandenberger}.

In spite of the great successes of inflation, it does not address one
of the most important aspects of the Big Bang cosmology: For any
equation of state obeying the strong energy condition $p>-\rho/3$,
regardless of the geometry (flat, open, closed) of the universe, the
scale factor of the universe in a Friedmann Robertson Walker (FRW)
metric vanishes at $t=0$, and the matter density diverges. In fact all
the curvature invariants, such as $R,~\Box R,...$, become singular.
This is the reason why it is called the {\it Big Bang singularity
problem}.

Although inflation requires $p<-\rho/3$, it does not alleviate the Big
Bang singularity problem, rather it pushes the singularity backwards
in time. Many authors have pondered on this debatable issue, whether
inflation is past eternal or not~\cite{Linde1,Borde}, and the
conclusion is that it is not, at least in the context of Einstein
gravity as long as the average expansion rate in the past is greater
than zero, i.e. $H_{av}>0$~\cite{Guth}. The fluctuations grow as the
universe approaches the singularity, and the standard {\it singularity
theorems} due to Hawking and Penrose hold, which inevitably leads to a
collapse in FRW geometry as long as the energy
density is positive~\cite{Hawking} (see also~\cite{Guth}).

There have been other attempts to circumvent the Big Bang singularity
problem due to anisotropic stresses, self regenerating universes
(during inflation), quantum cosmology, etc.\ (see~\cite{Linde}) but
none has successfully resolved the issue of space-like singularity,
especially in the context of a flat universe\footnote{In the context
of a closed universe where the curvature term acts as a ``source'' for
negative energy density in the Hubble equation, one can obtain
bouncing solutions~\cite{Starobinsky}, but this additionally requires
a phase of inflation, to address the ``flatness problem''. Although
this is a perfectly viable scenario, the motivations for a bounce is a
little bit lost, as it is also supposed to be an alternative to
inflation.}. Even string theory has yet to address this issue
comprehensively: Several toy model constructions have been attempted,
but mostly they encounter some pathologies, such as closed time-like
curves, quantum instabilities, presence of ghosts, negative-tension
branes, singular bounce, etc. (see for
instance~\cite{some})\footnote{ In~\cite{Burgess} the authors
considered several brane world models.  One such example is where a
probe anti-brane circles around a stack of source branes giving rise
to an induced $3+1$ dimensional metric. The bulk metric uausally has a
horizon and also singularities, but the induced metric can undergo a
non-singular bounce. One of the concerns of these models is the issue
of stability and lack of full control of the system.}.

In this paper we will rather seek a phenomenological solution, but
without any pathologies. We will seek a bouncing solution within $3+1$
dimensions  where the scale factor of a flat, homogeneous and isotropic
metric
\be
ds^2=-dt^2+a^2(t)\left[dx^2+dy^2+dz^2\right]
\ee
undergoes a {\it non-singular bounce}, a course of initial
contraction to a minimal radius, then a subsequent phase of expansion
to enter into the hot Big Bang era. 

We will advocate higher-derivative corrections to the Einstein-Hilbert
action that are both {\it ghost free} and {\it asymptotically free} to
find sensible non-singular bouncing solutions in the presence of a
fluid with {\it non-vanishing pressure}.  Earlier attempts to find
such an action were not very successful~\cite{Stelle} (see
also~\cite{Peter},\cite{Barton}). Below we briefly explain the kind of
actions that we consider and why we believe they can circumvent the
problems encountered in~\cite{Stelle},~\cite{Peter}, and others:

\begin{itemize}

\item{{\it Stringy motivation}:\\ 
The main theoretical motivation comes from string theory, which
suggests higher-derivative corrections to the Einstein-Hilbert action.
There such corrections appear already classically (\ie, at the tree
level), but we do not preclude theories where such corrections (or
strings themselves) appear at the loop level or {\it even
non-perturbatively} (e.g., as expected in the $1/N$ expansion in some
Yang-Mills theory~\cite{hooft}).  From string field theory~\cite{sft}
(either light-cone or covariant) the form of the higher-derivative
modification can be seen to be Gaussian: There are $e^\Box$ factors
appearing in all vertices (e.g., $(e^\Box\phi)^3$), which can be
transplanted to kinetic terms by field redefinitions ($\phi\to
e^{-\Box}\phi$). The nonperturbative gravity actions that we consider
will be inspired by such kinetic terms suggested by string theory.}

\item{{\it Ghost free action}:\\ Previously, non-singular bouncing
solutions have been obtained by introducing states carrying negative
energy compensating the positive energy contribution, by introducing
new fields or higher derivatives (see for instance~\cite{many}).  In
general, such theories are plagued by ghosts -- states that violate
unitarity. Alternatively, ghosts carry negative energy which
inevitably leads to vacuum instability~\cite{jim}.  Thus, one
condition we impose is the {\it absence of ghosts}.  Note that the
Gaussian factors of string theory do not introduce new states (ghosts
or otherwise) because they have neither poles nor zeroes at finite
momentum (which is also why they are allowed in field redefinitions).
}

\item{{\it Asymptotic freedom}:\\ 
This is required (in addition to renormalizability) in ordinary field
theory for phenomenological reasons (scaling in processes such as deep
inelastic scattering), as well as to avoid certain problems in the
nonperturbative definition of the theory.  (This can be seen, e.g., in
lattice quantum chromodynamics when defining a continuum limit.  In
the case of an effective theory for gravity, this condition takes the
form of ``asymptotic safety" \cite{Weinberg}, a more general statement
of a well-defined ultraviolet limit.)  In gravity this feature
provides a solution to the singularity problem; here we address the
Big Bang singularity.  (Black hole singularities were considered in
\cite{warren}.)  Asymptotic freedom implies that gravity becomes weak
at short distances or high energies, enabling pressure to counteract
the gravitational attraction, thereby preventing a collapse. Again the
Gaussian factors are more than enough to produce such behavior.  (The
fact that this fall-off is so much stronger than normal asymptotic
freedom is why string theory cannot describe parton behavior as a
model of hadrons for QCD.)}

\end{itemize}

We will find that if one wants to have both a ghost and asymptotically
free theory of gravity, one has little choice but to look into gravity
actions that are non-polynomial in derivatives. To see this let us
start by considering a simple generalization of the Einstein-Hilbert
action, the so called fourth-order gravity:
\be 
S=\int d^4x\ \sqrt{-g}F(R)\,, 
\ee 
with 
\be 
F(R)=R+c_0R^2+b_0C^2\,.
\label{R-square}
\ee
where $C$ refers to the Weyl tensor. This action, in general, is
asymptotically free. In fact, the improved ultraviolet behavior even
makes it renormalizable \cite{Stelle}. Unfortunately this theory also
has a (Weyl) ghost \cite{Peter} (see also \cite{Barton}), unless
$b_0=0$, in which case the nice property of asymptotic freedom is also
lost.  Similar conclusions can also be verified from the analysis done
by \cite{solganik} for more general actions involving the Ricci and
the Riemann tensors, when
$F=F(R,R_{\mu\nu}R^{\mu\nu},R_{\mu\nu\rho\s}R^{\mu\nu\rho\s})$.

In the most general case one can consider actions involving the scalar
curvature, the Ricci and Weyl tensors, as well as with arbitrary
higher derivatives on these curvatures.  If the number of derivatives
(including those in the curvatures) is finite, the kinetic operator
will be polynomial, and by the usual arguments will (1) contain a
single pole, (2) contain a ghost, or (3) not fall off at large momenta
faster than a single pole (i.e., for the case of gravity, not be
asymptotically free).  This argument can be applied separately to the
spin 0 and spin 2 pieces of the metric; only the spin 0 piece
contributes to cosmology, but the spin 2 piece contributes to black
holes.  In section~\ref{RBR} we discuss this for the spin 0 piece
(scale factor), which means the scalar curvature is the only
non-vanishing one (conformally flat space).

Therefore, in this paper we mainly focus on actions which contain
arbitrary powers of higher derivative terms: The simplest such action
is of the form
\be
F(R)=R+\sum_{n=0}^{\infty}c_{n}R^{n+2}\,.
\label{ordinary}
\ee
Unfortunately, even such actions cannot simultaneously be ghost and
asymptotically free, which is easily seen from its correspondence with
scalar-tensor/Brans-Dicke kind of theories (see for
instance~\cite{Wands}). Nevertheless, such theories are cosmologically
interesting alternatives to Einstein-Hilbert gravity, and can be
tested by the solar system constraints or through spinning objects
such as pulsars~\cite{Will}, or even
cosmologically~\cite{Andrew}. Such theories (or their scalar-tensor
analogues) have been studied in the context of
inflation~\cite{Starobinsky,Extended}, creation of the universe
through an instanton~\cite{Copeland}, understanding reheating after
inflation~\cite{Mendes}, and more recently, understanding the origin
of dark energy (see for instance~\cite{Us}). In section~\ref{RC} and
appendix~\ref{cond:a}, we review their dynamics in the context of
realizing a past asymptotically de Sitter universe and also point out
why it doesn't constitute a resolution of the Big Bang singularity,
although it may be relevant for inflation.

The next simplest non-perturbative action is of the form\footnote{Such
actions can also be motivated by symmetry considerations, such as
local (Weyl) scale transformation, and in section~\ref{EGA} we provide
such an example.}:
\be
F(R)=R+\sum_{n=0}^{\infty}c_{n}R\Box^{n}R
\label{quadratic}
\ee
We will show that such actions can indeed give rise to a ghost and
asymptotically free theory of gravity, and this is one of the key
results in the paper. Moreover, we will also obtain exact bouncing
cosmological solutions for such actions, thereby addressing the Big
Bang singularity problem, see section~\ref{BCAFG}. We note in passing
that (\ref{quadratic}) should be treated as an ``effective action'',
and thus the fact that it contains an infinite number of derivatives,
and is therefore non-local, is only to be expected.

We also study the Newtonian potential for the action given in
(\ref{quadratic}); we show that at large distances and at late
times contributions from the higher derivative terms become negligible
and the theory is effectively given by the Einstein-Hilbert
action. Therefore, the theory possesses Minkowski space as a low
energy vacuum solution, recovering the Newtonian potential with $1/r$
dependence, where $r$ is the length separation. On the other hand at
small distances the higher derivative terms play, more and more, an
important role, thereby indicating the importance of asymptotic
freedom near the bounce: see section~\ref{TFU}.

The paper is organized as follows: In section~\ref{RC} we mostly
review difficulties in obtaining a singularity free
universe. Partially for simplicity, we restrict ourselves to a
spatially flat universe. Specifically in the context of $F(R)$
theories we find that in the absence of ghosts, no satisfactory
singularity free evolution is possible, corroborating our earlier
arguments based on the lack of asymptotic freedom. We do note however
that a small class of actions that grow at most as $R^2$ as $R\ra
\infty$ can provide a past asymptotically de Sitter solution similar
to what was found previously in \cite{Starobinsky} for
(\ref{R-square}) and $b_0=0$.  In the following sections therefore we
focus on actions of the form (\ref{quadratic}).

In section~\ref{RBR} we explain when actions of the form
(\ref{quadratic}) can be both ghost and asymptotically free. We
also obtain the low energy Newtonian limit. In section~\ref{BCAFG}, we
obtain exact and approximate bouncing solutions, and in the next
section~\ref{TFU}, we show how a transition to normal FRW evolution
occurs before and after the bounce.
       

\section{$F(R)$ Cosmology}~\label{RC}

In the field theoretic framework a resolution to the Big Bang
singularity would imply a singularity free history from $t\ra -\infty$
till today. Here $t$ denotes the proper time. So the picture is that
the universe simply exists and one can go as far back in time as one
wants without encountering any singularity. A priori, three different
evolutions are possible:

\begin{description}

\item[I.\hspace{4mm}]{The universe expanded monotonically.}  

\item[II.\hspace{2mm}]{The universe first contracted and then
expanded (in other words, a ``bounce'' solution).}

\item[III.]{The universe undergoes periodic phases of contraction and
expansion, or a cyclic universe scenario.}

\end{description}

Since III requires multiple bounces and turnarounds, it also obviously
entails II, and therefore we will in this paper only focus on
determining whether I or II are possible.

In this section we study cosmological evolutions with $F(R)$
theories. We find that only a very restrictive class of such functions
can provide a singularity free history of our universe. To show this
we proceed in the following steps: First we explain the connection
between $F(R)$ theories given by (\ref{ordinary}) and scalar-tensor
theories \cite{magnano}.  Next, we study scalar-tensor theories and
find out for what kind of potentials one can have singularity free
cosmologies. Finally, we translate back the scalar-tensor action to
find which $F(R)$ actions can provide a singularity-free evolution
this way, but we also point out why they cannot address the Big Bang
singularity problem comprehensively.


\subsection{Conformal Transformation and Singularities}~\label{RN}

We start by noting that an action of the form (\ref{ordinary}) is
equivalent to a Brans-Dicke theory \cite{magnano} of the form
\be
S_{BD}=\int d^4x\ \sqrt{-g}[\Phi R-V(\Phi)]\,,
\ee
where $V(\Phi)$ is defined through a Legendre transform
\be
V(\Phi)=\Phi R(\Phi)-F(R(\Phi))\,,
\ee
with
\be
\Phi={dF\over dR}\equiv e^{\sqrt{\frac{2}{3}}\phi}\,,
\label{mapping}
\ee
defining $R=R(\Phi)$. Note that for a ghost free action $\Phi$ is
always positive; it was shown in \cite{solganik}, that when $F'(R)$ is
negative, one is plagued with ghosts. This fact then allows us to
perform a conformal transformation
\be g'_{\mu\nu}=\Phi
g_{\mu\nu}\equiv e^{\sqrt{\frac{2}{3}}\phi}g_{\mu\nu}\,, 
\ee 
so that the action is that of a scalar-tensor theory
\be
S_{ST}={M_p^2\over 2}\int d^4x\
\sqrt{-g'}\left[R'-(\p'\phi)^2-2{V(\phi)\over M_p^2}\right]\,,
\label{scalar-tensor}
\ee
where
\be
V(\phi)=\2e^{-\sqrt{\frac{2}{3}}\phi}R(\phi)-\2e^{-2\sqrt{\frac{2}{3}}\phi}
F(R(\phi))\,.
\label{potential}
\ee
Thus, at least classically, ghost-free $F(R)$ theories are physically
equivalent to the scalar-tensor theories given by
(\ref{scalar-tensor}). 

So, let us now ask when can one have a singularity free evolution in
the context of scalar-tensor theories. It is known already that such
theories with ordinary (non-ghost like) kinetic terms as in
(\ref{scalar-tensor}) do not permit a bouncing cosmology, as the sum
total contribution to the energy densities from the scalar and the
gravity sector remains positive. Therefore one does not obtain a
bouncing universe\footnote{There is an intriguing possibility that
although our universe may now be in a ghost free state, $F'(R)>0$, it
could have passed through an earlier phase when $F'(R)<0$. Can such a
``transient ghost'' degree of freedom cause a bounce?}, \ie,
situations summarized in II, III.

The only option left is I, \ie, to find a monotonically expanding
solution. Indeed, such solutions do exist within scalar-tensor
theories (see for instance~\cite{Starobinsky,Extended}). However, in
appendix~\ref{cond:a} we provide a general argument to show that
invariably such a scenario corresponds to an asymptotically de Sitter
universe in the past which is known to be geodesically
incomplete~\cite{Guth} and therefore fails to address the Big Bang
singularity in a comprehensive manner.  It may still be interesting
from the point of view of inflation. In appendix~\ref{cond:a}, we show
that: {\it only a subclass of $F(R)$ theory can provide such an
inflationary solution, only $F(R)$'s which grow at most as $R^2$ as
$R\ra \infty$.}


\section{Theories with $F(R,\Box R,\dots)$}~\label{RBR}

Since ghost-free $F(R)$ type actions do not satisfactorily resolve the
cosmological Big Bang singularity, we therefore look at more general
actions of type (\ref{quadratic}) including $\Box R$ type of terms in
the Lagrangian.


\subsection{Ghosts}

Although it is commonly believed that higher derivative theories
contain ghosts, this is not always true. For example, the $F(R)$
actions considered in the previous section do not contain any ghosts.
In Lorentz covariant gauges there appears to be a scalar ghost, along
with the usual ghostlike (negative-metric) time components of the
metric.  However, this ghost already appears in the usual
Einstein-Hilbert action: It is the scale factor of the metric.  It
disappears in the lightcone gauge, along with the longitudinal
components of the metric.  The addition of the $R^2$ term then
contributes a scalar mode of opposite sign, which is thus physical:
the right sign, and not a gauge artifact.  The same procedure would
not be possible with additional higher derivative contributions to the
propagator, which would give spin 0 or 2 with the wrong sign in any
gauge.  A simple way to analyze this effect for spin 0 modes is to
study the equivalent scalar-tensor theory where all the higher
derivative terms reside in the scalar sector, and one has to worry
about ghosts in the scalar sector only\footnote{Alternatively, one
can restrict oneself to conformal metrics, and just compute the
propagator for the conformal scale factor.}.

We start with the quadratic action given by (\ref{quadratic}). This
action is equivalent to a higher derivative scalar-tensor action,
given by
\be
S=\int d^4x\sqrt{-g}\left[\Phi R+\psi\sum_1^{\infty}c_i\Box^i\psi-
\{\psi(\Phi-1)-c_0\psi^{2}\}\right]\,.
\label{s-t}
\ee
The easiest way to see that (\ref{s-t}) corresponds to
(\ref{quadratic}) is by looking at the field equation for $\Phi$:
\be
{\de S\over \de \Phi}=0\Ra \psi=R\,. 
\ee
Then by substituting $\psi$ in (\ref{s-t}) one recovers
(\ref{quadratic}). We now perform a conformal transformation
$e_a{}^m=\Phi^{1/2}e'_a{}^m$, and note that $\Box\psi=\Box'\psi+{\cal
O}(\phi^2,\phi\psi,\psi^2)$ where we have defined
$\Phi=e^{\phi}\approx 1+\phi+{\cal O}(\phi^2)$.  Then up to quadratic
terms we find
\be
S\approx \int d^4x\ \sqrt{-g'}\left[R'+{3\over 2}\phi\Box'\phi +
\psi\sum_1^{\infty}c_i\Box^{'i}\psi-\{\psi\phi-c_0\psi^{2}\}\right]\,.
\ee 
To look for ghosts we have to find the propagators. The field
equations for $\phi$ and $\psi$ are given by:
\begin{eqnarray}
\label{phi-eqn}
\psi&=&3\Box\phi\,, \\
\phi&=&2\left[\sum_1^{\infty}c_i\Box^{i}\psi+c_0\psi\right]\,.
\label{phi0-eqn}
\end{eqnarray}
Substituting $\psi$ from (\ref{phi-eqn}) in (\ref{phi0-eqn})
we find:
\be
\left(1-6\sum_0^{\infty}c_i\Box^{i+1}\right)\phi\equiv \Ga(\Box)\phi=0\,.
\label{final-eqn}
\ee
From (\ref{final-eqn}) we can easily read off the scalar propagator to
be:
\be
G(p^2)\sim {1\over \Ga(-p^2)}\,.
\ee
Let us consider the contrasting cases.


\begin{itemize}

\item {\it Examples of  ghosts}:

First we assume that $\Ga$ is a finite power series. In this case one
can always write $\Ga$ as:
\be
\Ga(-p^2)\sim (p^2+m_1^2)(p^2+m_2^2)\dots(p^2+m_n^2)\,.
\label{poly-prop}
\ee
In order for the theory to be non-tachyonic, all the $m_i^2$ have to
be positive and real. Moreover, if there are at least two discrete
single poles (say $m_1\neq m_2$), then at least one of them is ghost
like (one of the residues has to be negative).
A double pole can be represented as the convergence of two simple
poles with opposite residues.  Similar arguments follow for higher
order poles.

The above argument can be generalized to any continuous
$\Ga(-p^2)$. Suppose $ \Ga(-p^2)$ has at least two distinct
zeroes. Say $p^2=-m_1^2$ and $p^2=-m_2^2$ are two adjacent ones, with
$m_1^2<m_2^2$. Then it is of the form
\be
\Ga(-p^2)=(p^2+m_1^2)(p^2+m_2^2)f(-p^2)\,.
\ee
Since the zeroes considered above are adjacent there are no more
zeroes in $f(-p^2)$ between $-m_2^2<p^2<-m_1^2$. Thus the sign of
$f(-p^2)$ has not changed in this range. It then follows that the
residue at $p^2=-m_1^2$ and $p^2=-m_2^2$ have different
signs. Therefore again there is a ghost.


\item {\it Requirement for a ghost free case}:

For a polynomial inverse propagator (\ref{poly-prop}) the only case
when the propagator is {\it ghost free} is
\be
\Ga(\Box)=\Box-m_0^2\,,
\ee
which corresponds to the $R^2$ type of action that we mentioned in the
previous section (also, see
appendices~\ref{cond:a},\ref{Bouncepert}). More generally, one can at
most have a single zero in $\Ga(-p^2)$.

As we argued before, stringy redefinitions are not expected to
introduce new states, and hence we will only consider the case when
$\Ga(-p^2)$ has, in fact, no zeroes.  Partially for simplicity, we
also assume it to be {\it analytic} in the entire complex plane
(except perhaps at $\infty$). In this case one can write $ \Ga(-p^2)$
as:
\be
\Ga(-p^2)=e^{\ga(-p^2)}\,.
\ee
where $\ga(-p^2)$ is any analytic function.

\end{itemize}


\subsection{Newtonian Potential and Asymptotic Freedom}\label{AF}

To derive the Newtonian potential one has to linearize the
``generalized Einstein's field equations'' corresponding to the action
given in (\ref{quadratic}) (see appendix~\ref{Newt} for details). In
particular, the trace of the linearized field equations coupled to a
point source reads:
\be 
-\2 \Box\left(1-6\sum_0^{\infty}c_i\Box^{i+2}\right)\hb =
-\2\Box\Ga(\Box)\hb\sim \de(\vec{r})\,. 
\ee 
Since we are only interested in cosmology, we only need to look at the
potential for the scale factor or the spin $0$ piece, which is
captured by the trace $\hb$ of the linearized metric. Note that in the
absence of any modification the massless propagator of the graviton
can be captured alone by the $\Box$ in front of $\Gamma(\Box)$.

We now further make the assumption that the fields are varying slowly
with time, so that $\Box\longrightarrow \nabla^2$, and after some
massaging we obtain:
\be 
\hb(r)\sim \int d^3p\
{e^{i\vec{p}.\vec{r}}\over p^2\Ga(-p^2)} \sim
{-i\over r}\int_{-\infty}^{\infty} dp {e^{-\ga(-p^2)}\over p}e^{ipr}\,.
\label{newtpot}
\ee
Note that the $1/r$ part arises mainly from the massless mode of the
graviton. Moreover this is an important result of our paper as it
tells us how higher curvature gravity in general modifies the
propagator, and therefore the nature of gravity, as one spans very
large to very small scales.  For example it is clear that:


\begin{itemize}

\item{\it Any modification to the ``Newtonian propagator'' $\Ga$ that falls
faster than $1/p$ can give rise to an asymptotically free theory of
gravity. In particular, as long as $\ga(-p^2)$ grows with large $p^2$,
we will not only have a {\it ghost free} theory, but also asymptotic
freedom. }
\end{itemize}
 
For simplicity, let us restrict our attention to the case when $\ga$
is a finite power series in $w\equiv -p^2$, and that there is a single
scale $m$ where the higher derivative effects become important:
\be
\ga(w)=\sum_{i=1}^N k_i\left(-w\over m^2\right)^i\,,\quad k_i\sim {\cal
O}(1)
\label{gamma}
\ee 
Asymptotic freedom is guaranteed provided $k_N>0$. From
Eqs.~(\ref{newtpot}) and (\ref{gamma}), it also follows that we
recover Newtonian gravity in the large $r$ limit, provided
$m>10^{-3}$~eV, since we have tested gravity only at distances above
the mm.\ scale, which corresponds to $m\sim 10^{-3}$~eV.


\subsection{Examples of Ghost and Asymptotically Free Theories}\label{EGA}

For the purpose of illustration, let us choose the simplest example:
\be
\ga(w)=-w \Ra \Ga(w)=e^{-w}\,.
\label{exponential}
\ee
By construction this does not have any poles and is therefore ghost
free\footnote{Such exponential propagators also appear in field
theory duals of ``discretised'' string theory, for instance see
\cite{marc}.  This type of modification to gravity has also been considered in
\cite{expo}.}.

Let us first find the action that corresponds to
(\ref{exponential}). From (\ref{final-eqn}), we can immediately match
the coefficients by equating:
\be
\left(1-6\sum_0^{\infty}c_i\Box^{i+1}\right)=e^{-\Box}\,,
\ee
which corresponds to
\be
c_i=-{1\over 6}{(-)^{i+1}\over (i+1)!}\,.
\ee
For the above choice the form of $F(R)$ can be computed as:
\be
 F(R)=R-{1\over 6}R\sum_{i=0}^{\infty}{(-)^{i+1}\over (i+1)!}
\Box^iR=R-{1\over 6}R\left({e^{-\Box}-1\over \Box}\right)R\,.
\label{exp-F}
\ee
Next, let's look at the Newtonian limit. The Newtonian potential is
given by
\be
\hb(r)={i\over r}\int_{-\infty}^{\infty} 
dp {e^{-p^2}\over p}e^{ipr}={\pi\over r}\mt{erf}\left({r\over r_{0}}\right)\,.
\ee
where $r_{0}$ is the ultraviolet cut-off, determined by the energy
scale, governing the graviton exchange.

Since $\lim_{~r\ra 0}\mt{erf}(r)\sim r$ and $\lim_{~r\ra
\infty}\mt{erf}(r)= 1$, we naturally obtain the kind of Newtonian
potential we are seeking at large distances, \ie,
\be 
\lim_{~r\ra\infty}\hb(r)\sim -{1\over r}\,,
\ee
while for small $r$ we are looking for an asymptotically free
theory, with no singularity: In this case
\be
\lim_{r\ra0}\hb(r)\sim \frac{1}{r_{0}}=\mt{const.}\,.
\label{newtonian}
\ee
This illustrates the weakness of gravity at short distances when
$r\rightarrow r_{0}$.

Let us now illustrate how actions like (\ref{quadratic}) can also
arise from symmetry consderations. For example, one realizes that a
slight modification of the action corresponding to (\ref{exponential})
leads to a form that is relatively simple in conformally flat spaces.
Consider introducing a scale factor into the action by a local (Weyl)
scale transformation $g_{mn}\to\phi^2 g_{mn}$.  This scalar $\phi$
itself has weight 1 under a scale transformation.  The combination
$(\Box+\frac16R)\phi$ also transforms covariantly, with weight 3.
Then we can write
\be 
S=\int d^4 x\ \sqrt{-g}\ \phi^3
e^{\phi^{-2}(\Box+\frac16R)}\phi^{-2}(\Box+\frac16R)\phi \ .  
\ee 
In
the local scale gauge $\phi=1$, this becomes: 
\be 
S \to \int d^4 x\
\sqrt{-g}\ e^{(\Box+\frac16R)}\frac{1}{6}R,
\label{non-linear}
\ee
while for conformally flat spaces we can set the metric to be flat
($\phi$ itself being the scale factor) so
\be
 S \to \int d^4 x\ \phi^3
e^{\phi^{-2}\Box}\phi^{-2}\Box\phi\ , 
\ee
where $\Box$ is now the flat one. It is easy to check that up to
quadratic (in $R$) terms, (\ref{non-linear}) corresponds exactly to
(\ref{exp-F}).


\section{Bouncing Cosmology from Asymptotically Free Gravity}~\label{BCAFG}

In the previous section we saw how nonperturbative gravity
theories can encompass asymptotic freedom, which provides hope for
resolving various singularities we encounter in the general theory of
relativity.  Here our main focus will be on the initial cosmological
singularity. Although at first glance it may seem very difficult to
understand the dynamics involving higher derivative actions such as
(\ref{quadratic}), we will in fact be able to explicitly show
approximate and exact bouncing solutions for a wide class of higher
derivative gravity theory.

First, we discuss why intuitively we expect to obtain bouncing
solutions in asymptotically free theories. Second, we provide exact
bouncing solutions for ghost and asymptotically free non-perturbative
gravity actions described in (\ref{quadratic}) in the presence of a
cosmological constant. Although phenomenologically less appealing, it
gives insight into the nature of gravity near the bounce.  In the next
section, we show how to obtain approximate bounce solutions in the
absence of a cosmological constant, as well. We also highlight how to
match the bouncing solutions to the standard radiation and matter
dominated epoch.


\subsection{Intuitive Picture}\label{ITA}

Intuitively, the reason we expect to have a bounce in theories with
higher curvature gravity is because, although they do not have ghosts,
gravity itself becomes weak at small scales. In the usual FRW
cosmology, it is the attractive force of gravity which makes the
universe contract, eventually to a singularity. However, now in the
absence of strong gravity at short distances (or high energy
densities) spacetime contracts but the internal pressure of matter can
resist, therefore causing the universe to bounce from the contracting
phase into an expanding one.

In the Newtonian cosmology, the Hubble equation can be derived from
the energy conservation equation of particle mechanics, \ie, $K+V=E$,
\be
\2m\dot{r}^2+G_N{4\pi\over 3}r^3\rho V(r)=E\,.
\ee
Substituting $V(r)=-1/r$, we recover the analog of the Hubble equation:
\be
H^2\equiv {\dot{r}^2\over r^2}={M_p^2\over 3}\left(
\rho + {6E\over M_p^2m r^2}\right)\,,
\ee
where the second term on the right corresponds to spatial curvature.

Specializing to the case of a flat space ($E=0$) and considering a
general Newtonian potential arising from the higher curvature gravity
action derived in previous section, we find
\be \label{NewtH}
H^2\equiv {\dot{r}^2\over r^2}=-{M_p^2\over 3}\rho rV(r)\,.
\ee
In an asymptotically free theory, at short distances the potential
tends to a {\it constant}, so the right hand side of (\ref{NewtH})
goes to zero, suggesting $H\ra 0$, a condition signaling a bounce.

One may wonder that perhaps one hits the singularity before the bounce
can take place, because the gravitational energy vanishes only at
$r=0$. However, note that we have not included effects of pressure in
(\ref{NewtH}), which ensures that the bounce occurs at a finite scale
factor. Unfortunately, Newtonian cosmology cannot capture the essence
of pressure, but in the following subsections we will explicitly
realize this intuitive picture in a large class of higher derivative
actions.


\subsection{Hyperbolic Cosine Bounces}

To understand how one can deal with non-perturbative actions
and equations of motion, let us look at the trace of the generalized
Einstein equation of motion, given in appendix \ref{Newt}, see
(\ref{G}):
\be
\ti{G}\equiv G+ 6c_0\Box R + 
\sum_{m=1}^{\infty}c_m\left(6\Box^{m+1}R+
2\sum_{n=1}^m\Box^{m-n}R\Box^nR\right)=-4\La\,,
\label{Gtrace}
\ee 
where $\La$ refers to the cosmological constant. Note that the trace
equation involves $\Box$'s acting on $R$. Thus the trick is to find a
scale factor $a(t)$ such that $R(t)$ is a linear combination of a set of functions $H_i(t)$ with the property  that the
$\Box$ maps the functions $H_i(t)$ onto themselves:
\be
\Box H_i=M_{ij}H_j\,,
\label{recursion}
\ee
where $M_{ij}$ is a constant matrix and the index $i$ runs to some
finite value. We can now find exact solutions, because all the terms
in (\ref{Gtrace}) are at most quadratic in $H_i$ and one has to only
ensure that the overall coefficients of $H_i H_j$
vanish\footnote{This is not strictly necessary as sometimes they can
be equated to matter sources with similar time dependences.}.

A simple trial bounce solution is be given by\footnote{There exist
other ansatze, such as $a(t)\propto e^{t^2}$, that can also provide
non-singular bounce. For the purpose of illustration we shall stick to
(\ref{cosh}).}:
\be
a(t)=a_0\cosh\left(\sqw t\right)\,.
\label{cosh}
\ee
One finds the following:
\begin{eqnarray}
H&=&\sqw\tanh\left(\sqw t\right)\,, \\
R&=&3w\left[2-\mt{sech}^2
\left(\sqw t\right)\right]\,, \\
\Box R &=&-(\ddot{R}+3H\dot{R})=-3w^2\left[\mt{sech}^2\left(\sqw t\right)
\right]=w[R-6w]\,.
\end{eqnarray}
{\it Naively one would have expected $\Box R$ to contain terms
proportional to $\mt{sech}^4$, but the contribution to such a term
coming from the two pieces (the double derivative and the connection
terms) cancel leaving only the $\mt{sech}^2$ piece}, which is exactly
what we wanted. Proceeding in a similar manner one finds
\be
\Box^n R=w^{n-1}\Box R=-3w^{n+1}\mt{sech}^2\left(\sqw t\right),\ n\neq 0\,.
\label{box-n}
\ee
In this example:
\be
H_0=1\,,~~~~~H_1=\mt{sech}^2\left(\sqw t\right)\,.
\ee

Before we delve into bounces in non-perturbative gravity, let us
briefly mention what happens in the perturbative cases. Clearly, the
prescription described above can be applied to the finite higher
curvature theories as well to obtain hyperbolic cosine
bounces. Albeit, one can show (see appendix~\ref{Bouncepert}) that
they invariably involve negative energy density, in congruence with
our earlier arguments. They either have ghosts in the gravity sector,
or require the presence of ghost-like radiation with negative energy
density.

This also implies that we check that the radiation energy density
corresponding to the bounce solution remains positive to ensure that
the bounce is caused really by the higher derivative effects.  This
means that we need to study the $\ti{G}_{00}$ equation, which is
really the generalization of the Hubble equation in Einstein
gravity. For an ansatz like (\ref{cosh}) where the metric only depends
on time, one finds (from (\ref{generalized})):
\be
\ti{G}_{00}=F_0R_{00}+{F\over 2}-F_{0;\ 00}-\Box F_0-
\2\sum_{n=1}^{\infty}F_n\Box^n R-{3\over 2}
\sum_{n=1}^{\infty}\dot{F}_n\dot{(\Box^{n-1}R)}\,.
\label{tildeG00}
\ee
After some some algebra we obtain 
\be
\ti{G}_{00}=L_0+L_2\mt{sech}^2\left(\sqw t\right)+
L_4\mt{sech}^4\left(\sqw t\right)+L_6\mt{sech}^6\left(\sqw t\right)\,,
\ee
where one can conveniently express all the coefficients in terms of
the kinetic operator $\Ga(\Box)$:
\begin{eqnarray}
\label{L0}
L_0(w)&=& {w\over 2}\,, \\
L_2(w) &=&{9w^2\over 2}\left[{1-\Ga(w)-4c_0w\over w}\right]-{3w\over 2}\,, \\
\label{L4}
L_4(w) &=& {9w^2\over 2}\left[2c_0+{5\over 6}\Ga'(w)\right]\,, \\
L_6(w) &=&{9w^2\over 2}\left[{\Ga(w)-1-\Ga'(w) w\over w}\right]\,.
\label{L6}
\end{eqnarray}
Now, we wish to solve the $00$ component of the generalized Einstein
equation, (\ref{generalized}), which reads:
\begin{eqnarray}
\ti{G}_{00} &=& L_0+L_2\mt{sech}^2\left(\sqw t\right)+
L_4\mt{sech}^4\left(\sqw t\right)+L_6\mt{sech}^6\left(\sqw t\right)\, 
\nonumber \\
& =& T_{00}=\3\left[\La +\rho_0\mt{sech}^4\left(\sqw t\right)\right]\,.
\label{exact-eqn}
\end{eqnarray}
where in the $00$ component of the energy momentum tensor we are
compelled to include the cosmological constant term to match
$L_0$, besides adding the radiation component. The radiation component
must fall as $\mt{sech}^4$ following $\rho\sim a^{-4}$. The above
(\ref{exact-eqn}) implies that there exists a solution if
$L_2(w)=L_6(w)=0$ or $1-\Ga(w)+\Ga'(w) w =0$, and
$1-\Ga(w)-4c_0w-\3=0$, or equivalently,
\be
\Ga'(w)={\Ga(w)-\Ga(0)\over w} = \langle\Ga'(w)\rangle = {2\over 3}\Ga'(0)-{1\over 3w}
\label{constraints}
\ee
(where $\langle\ \rangle$ is the average between 0 and $w$).
For the above solution, equating the $\mt{sech}^4$ term and the
cosmological constant, we further find:
\be
\rho_0={3w\over 4}(2w\Ga'(w)-1)\,~~~~~~\mx{ and }~~~~~~\La= {3w\over 2}\,.
\ee
To be a ghost free solution, we require: $\rho_0>0$
\be
\Rightarrow~~2w\Ga'(w)-1\geq 0.
\label{radiation}
\ee
Thus, any kinetic operator $\Ga$ satisfying Eqs.~(\ref{constraints}) and
(\ref{radiation}) provides us with a ghost-free cosine hyperbolic bounce solution. We note that such cosine hyperbolic bouncing metrics and their generalizations (some of which also satisfy (\ref{recursion})) were studied in~\cite{mcinnes}. In particular
it was shown that although such metrics violate the dominant energy condition, causality is preserved and, as expected in bouncing universes, they can address the horizon problem.


\subsection{An Example}

In this subsection we provide a specific example which serves as an
existential proof for a ghost free bounce. We argue that the above
constraints, (\ref{constraints}) and (\ref{radiation}), can be
satisfied for a wide range of $\Ga(w)$. Let us consider the following:
\be
\ga(w)=k_1w-k_2w^2+k_4w^4\,,
\label{quartic}
\ee
where $k_{1},~k_2,~k_4$ are constants. Let us suppose that there
exists a solution to (\ref{constraints}) at $w=w_0$ and 
\be
\Ga(w_0)=\Ga_0
\label{point}
\ee
Then the $L_2$ and $L_6$ constraints, along with (\ref{point}), fix all the
constants:
\begin{eqnarray}
\nonumber
k_1 &= & {1\over 2w_0}(3\Ga_0-2)\,.\\ \nonumber
k_2 &=&{1\over w_0^2}\left[{9\Ga_0^2-4\Ga_0-2\over 4\Ga_0}-2\ln \Ga_0
\right]\,\\
k_4 &=&{1\over w_0^4}\left[{3\Ga_0^2-2\over 4\Ga_0}-\ln \Ga_0\right]\,.
\label{k-coefficients}
\end{eqnarray}

We compute the radiation energy density and we find:
\be
\rho_0={3w_0\over 2}(2\Ga_0-3)\geq 0\im \Ga_0>{3\over 2}\,,
\ee
indicating that $\Ga(w)$ of the form (\ref{quartic}), with
coefficients given by (\ref{k-coefficients}), indeed gives rise to
a ghost free bouncing solution, provided $\Ga_0>3/2$. $w_0$ remains a
free parameter.

The cosmological constant corresponding to this solution is given by
$\La={3w_0\over 2}$. We notice that we seem to require a large
cosmological constant in order for the exact bounce solution to exist,
just as in the $R^2$ example shown in appendix
\ref{Bouncepert}. However, in the next section we will argue that a
large non-vanishing cosmological constant is rather an artifact of our
ansatz, \ie, $a(t)\propto \mt{cosh}\ t$, but is not a requirement for
the bounce itself.


\section{Transition to FRW Universe}\label{TFU} 

It is clear that for large $t$ the scale factor has a de Sitter solution:
\be
a(t)\ra {a_0\over 2}e^{\sqw t}\,.
\label{inflation}
\ee
This corresponds to an inflation scenario that has no graceful
exit. On the other hand, it seems a general feature of the exact
``cosh'' bounce solutions that they require a cosmological constant
(typically $\sim {\cal O}(M_p^4)$) for their existence.  These two
problems are related: Note as $t\ra \infty$ all the higher curvature
terms vanish:
\be
\ti{G}_{00}^{HD}\equiv\sum_{n=0}^{\infty}\ti{G}_{00}^n\ra 
\mt{sech}^2\left(\sqw t\right)\ra 0\,,
\label{hc-energy}
\ee
and we are just left with Einstein's theory of gravity. The
cosmological constant is precisely required to sustain the inflation
in this phase (\ref{inflation}), and thereby ensures that the
asymptotics of our ansatz is consistent.

What is important is to realize that the cosmological constant is
mainly required for the late-time consistency, see (\ref{inflation}),
for our ansatz (\ref{cosh}). It is certainly not necessary for having
a bounce. In fact, we argue that its absence solves the graceful exit
problem.


\subsection{Approximate Bounce Solutions}

We have seen how for large $t$ all the higher curvature contributions
in (\ref{tildeG00}) vanish, leaving only the lowest order,
Einstein-Hilbert term.  In contrast, exactly at the bounce the
contribution from the Einstein-Hilbert action is zero, while those
coming from the higher order terms are finite. Thus, to understand the
dynamics it seems natural to divide the evolution into two distinct
regimes:

\begin{enumerate}

\item{near the bounce when the higher curvature terms dominate
the evolution}

\item{away from (both before and after) the bounce when the higher
order terms can be ignored as compared to ordinary gravity and we have
normal FRW evolution}

\end{enumerate}

In the next subsection we will study the transitions. Here we focus on
how to obtain the approximate bounce solutions, \ie, phase 2.

As compared to (\ref{exact-eqn}), the approximate equation that we
now wish to solve contains only the higher derivative terms,
\be
\ti{G}^{HD}_{00}=T_{00}
=\rho_0\mt{sech}^4\left(\sqw t\right)\,.
\label{G00}
\ee
We neglected the contribution coming from ordinary Einstein gravity
and assumed $\La=0$. In the next subsection we will check that it is
indeed consistent to neglect the Einstein-Hilbert terms near the
bounce. Now, it is clear from~(\ref{L0}-\ref{L6}) that
$\ti{G}^{HD}_{00}$ is of the form:
\be \ti{G}^{HD}_{00}=L^{HD}_2\mt{sech}^2\left(\sqw t\right)+
L_4\mt{sech}^4\left(\sqw t\right)+L_6\mt{sech}^6\left(\sqw t\right)\,,
\ee 
where $L_4$ and $L_6$ are given as before by~(\ref{L4},~\ref{L6}),
while
\be
L^{HD}_2={9w^2\over 2}\left[{1-\Ga(w)-4c_0w\over w}\right]\,.
\ee
(\ref{G00}) then implies $L^{HD}_2(w)=L_6(w)=0$, or
$1-\Ga(w)+\Ga'(w) w =0\mx{ and }1-\Ga(w)-4c_0w=0$, or equivalently,
\be
\Ga'(w)={\Ga(w)-\Ga(0)\over w} = \langle\Ga'(w)\rangle = {2\over 3}\Ga'(0)\,.
\label{constraints-1}
\ee
Moreover, equating $\ti{G}^{HD}_{00}$ with the stress energy
tensor, we find:
\be
\rho_0={3w^2\over 2}\Ga'(w)\,.
\ee
%

\FIGURE[!h]{
\includegraphics[scale=0.6,angle=0]{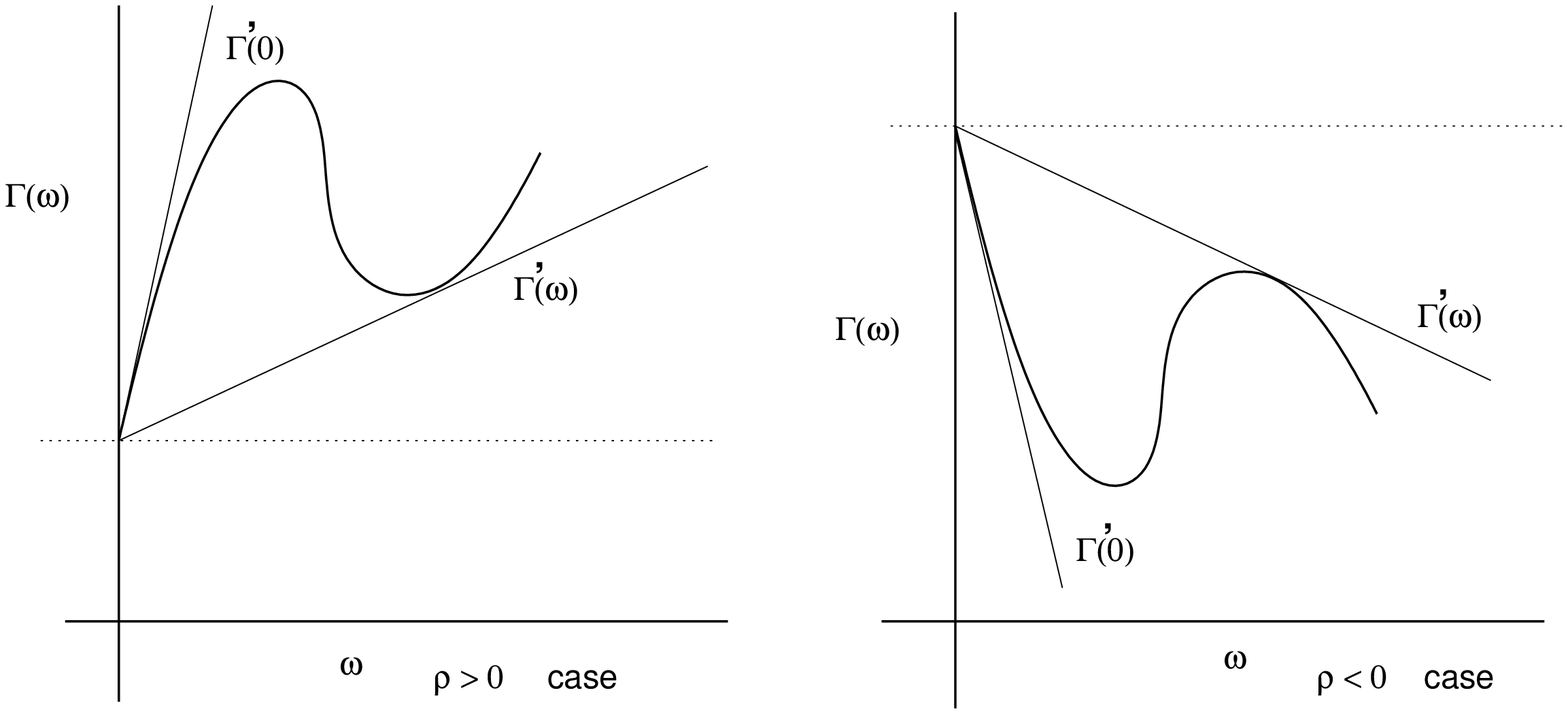}
%
\caption{\small Geometrical Solutions to the ``Bounce Constraints''}
\label{figps1}
}


The constraints (\ref{constraints-1}) have simple geometrical
interpretations. We can understand the solutions just by looking at
the curve $\Ga(w)$ (see figure~\ref{figps1}). The function $\Ga(w)$
starts at $\Ga(0)=1$, and possibly passes through a finite number of
minima and maxima on its way toward $w\ra \infty$. The solution to the
$L_6$ constraint is given by the points at which tangents from the
point $(w=0,\Ga(0)=1)$ touch the curve $\Ga(w)$. The $L^{HD}_2$
constraint then simply says that the slope of this tangent has to be
equal to two-thirds the slope of the curve at $(w=0,\Ga(0)=1)$.

Finally, the radiation energy density is also proportional to the
slope of this tangent, so demanding a positive energy density solution
simply means having a positive tangential slope. We note in passing
that these constraints are very similar to the ones that we obtained
for the exact bounce case.

Again, as an existential proof one can check that for 
\be
k_1={3\over 2}\ ;\ k_2={5\over 2}-2\ln 2\,~~~~~\mx{ and }~~~~ k_4=1-\ln 2\,,
\ee
in (\ref{quartic}), there exists a solution to the constraints at $w=1$
and $\Ga(1)=2$, with positive radiation energy density $\rho_0=3/2$.
 
The pictorial understanding of the ``bounce constraints'' makes it
easy to see what kind of $\Ga(w)$ can give rise to a bounce. However,
we are still unable to clearly identify a connection of these
constraints to the ``necessity'' or ``sufficiency'' of asymptotic
freedom. For example, it is clear that $\Ga(w)$ corresponding to just
the simple exponential as discussed in section~\ref{EGA} does not
satisfy the bounce equations. The most likely reason is that we only
looked at a very specific bounce solution, namely the hyperbolic
cosine bounce, and therefore surely are missing more complicated
bounces which may be present in some of the asymptotically free
theories. We reserve a more detailed study of these issues to future
research.


\subsection{Exiting the Bounce}~\label{ETB}

In this subsection we proceed as follows: First, we show why one can
trust the approximate bounce solution that we have obtained in the
previous subsection. We also estimate when the bounce solution breaks
down and one expects a normal FRW universe to emerge. Second, we check
that indeed away from the bounce the usual FRW cosmology is a good
description and higher derivatives can be ignored.  Finally, we obtain
the full evolution of the universe by matching the scale factor and
its derivatives at the boundary of the different regimes in a
straightforward manner. For simplicity in the following we consider
the action to be basically governed by a single scale as in
(\ref{gamma}) and therefore, unless there is some fine-tuning, one
expects $w\sim m^2$. Let us now investigate the different regimes of
the evolution.

As we argued earlier, (\ref{cosh}) is not an exact solution but it is
still a good approximate solution at sufficiently high energy
densities. One way to see this is to consider the left-hand side of
(\ref{G00}) close to the bounce. In this regime, the higher order
terms dominate over the Einstein-Hilbert term in the evolution
equation (\ref{G00}): $G_{00}={3w\over 2}(1-\mt{sech}^2\sqw t)$ while
$\ti{G}^{HD}_{00}={3w^2\over 2}\Ga'\mt{sech}^4\sqw t$. For example,
exactly at the bounce ($t=0$) $G_{00}$ vanishes while
$\ti{G}^{HD}_{00}\sim w^2\Ga'\sim w$ is finite.

However, as $t$ becomes large, all the higher order corrections get
exponentially suppressed:
\be
\mt{sech}^4\left(\sqw t\right)\sim e^{-4\sqw t}\,,
\ee
while the zeroth order term coming from the ordinary Einstein-Hilbert
action is not: There is a constant piece
\be
G_{00}={3w\over 2}\left[1-\mt{sech}^2\left(\sqw t\right)\right]
\st{|t|\ra \infty}{\longrightarrow} {3w\over 2}\,.
\ee
Thus, the bounce solution breaks down approximately when
\be
G_{00}=\ti{G}^{HD}_{00}\Ra \sqw t_0=\pm\cosh^{-1}\sqrt{2w\Ga'\over 
\sqrt{1+4w\Ga'}-1}\,,
\label{t0exact}
\ee
and the Einstein-Hilbert term starts governing the evolution from here
on. Since ${w\over 2}\Ga'\sim {\cal O}(1)$, we have:
\be
\sqw t_0\sim {\cal O}(1)\,\mx{ or }t_0\sim \sqrt{1\over w}\sim m^{-1}
\label{t0}
\ee 

Let us now check that indeed there exists late-time attractor
solutions where higher derivative terms can be ignored and one has FRW
cosmology. The transition from the bounce should lead to a radiation
dominated epoch before and after the bounce.  During these phases, we
claim we have ordinary gravity coupled to an ideal gas of
matter/radiation fluid satisfying
\be \rho=\om p\mx{ and }\dot{\rho}+3Hp=0\Ra \rho=\rho_0 \left({a\over
a_{0}^{\pm}}\right)^{-3(1+\om)}\,,  
\ee 
where $-$ and $+$ refer to the pre and post bounce solutions
respectively.  Therefore, at large times the solution reads:
\be
a(t)=a_{0}^{\pm}\left({t\over \pm t_0}\right)^{2/3(1+\om)}
\equiv a_{0}^{\pm}\left({t\over \pm t_0}\right)^{p}\,.
\label{pre-bounce}
\ee
To see that this is a late-time attractor solution, let us
compare $G_{00}$ with $\ti{G}^{HD}_{00}$.  For the power law
solutions, we obtain:
\begin{eqnarray}
\label{R-term}
 G_{00}\sim {1\over t^2}\,, \\
\label{box-R}
\ti{G}^{0}_{00} \sim {c_0\over t^4}\sim {1\over m^2 t^4}\,.
\end{eqnarray}
Thus, for large times $|t|>>m^{-1}$, the higher order terms are
negligible, and indeed (\ref{pre-bounce}) is a late-time attractor
solution. However, when we reach $|t|\leq m^{-1}$, the
$\ti{G}^{0}_{00}$ term (and in general $\ti{G}^{HD}_{00}$) starts to
dominate over the the Einstein-Hilbert term. The solution
(\ref{pre-bounce}) is no longer valid. At this point we expect a
transition to the approximate bounce solution given by (\ref{cosh}).
By matching the scale factor and its derivatives at the transition, we
find:
\be
a_0^{\pm}=\cosh\left(\sqw t_0\right)\,,
\ee
with $t_0$  given by 
\be
\sqw t_0\tanh\left(\sqw t_0\right)=p\,.
\label{t00}
\ee
In particular, for radiation $p=1/2$ and $\sqw t_0\approx 0.77$,
consistent with (\ref{t0}).

To summarize, the full evolution is given by 
\be
a(t)=\left\{ \begin{array}{ll}
\cosh\left(\sqw t_0\right)({t\over -t_0})^{p} &\mx{ for }t<-t_0 \\ 
\cosh\left(\sqw t \right)&\mx{ for }t_0>t>-t_0\\
\cosh\left(\sqw t_0\right)({t\over t_0})^{p}&\mx{ for }t>t_0
\end{array}\right.
\ee
Note the scale $\sqw$ is governed by the higher curvature action,
while $t_0$ is given by (\ref{t0exact}).  This completes our discussion on
exiting an approximate bounce solution to the normal FRW universe.

\section{Future Directions}

In this work we have shown how non-perturbative effective actions of
gravity are not only natural from the string theory perspective but
also are able to realize a ghost and asymptotically free theory of
gravity. Further, from our analysis it is clear that one can apply
different techniques (like the Newtonian approximation, special
ansatze, etc.) to understand the dynamics of the full non-perturbative
theory. In particular, we obtained bouncing solutions and Newtonian
potentials, which enabled us to argue how at short distances/times
these theories can address singularities like the Big Bang
singularity, while approximating ordinary General Relativity and FRW
cosmology for large distances and times.

Although we could only understand the transition from bounce to FRW
cosmology in an approximate way, nevertheless this supports our view
that bouncing cosmology can exist in a {\it flat} geometry without any
requirements such as presence of ghosts or tachyonic states.

In the light of observational data from WMAP, another important
question would be the evolution of cosmological perturbations through
the bounce, as well as their spectrum. It would also be interesting to
investigate whether such a bounce leaves any distinctive signature in
the production of gravitational waves, etc. Further note the fact that
we could perform an exact analysis of non-perturbative actions provides
us hope of trying to address other singularities, such as one
encounters in the black hole physics. 

In brief, there are several interesting and important questions that
we can now ask in the context of non-perturbative gravity.

\section*{Acknowledgments}

The authors are thankful to Robert Brandenberger and Cliff Burgess for
helpful discussions, and particularly to Marc Grisaru for
collaborating during the initial stages of the project. A.M. would
like to thank the Aspen Center for Physics for hospitality during the
final stages of this work.


\appendix


\section{Condition for a Singularity Free Universe in $F(R)$ Cosmology}\label{cond:a}

We start by looking at scalar-tensor theories (\ref{scalar-tensor}),
which can give rise to a monotonically expanding singularity free
evolution, namely condition I.  Note that the curvature scalar is
given in terms of the scale factor,
\be
R=6\left(\frac{\ddot{a}}{a}+\left(\frac{\dot{a}}{a}\right)^2\right)\,.
\label{curvature}
\ee
From (\ref{curvature}) it is clear that we will have a singularity
in the past if $a = 0$ at a finite time.  This can be seen very easily: Say $a=0$ at
$t=0$, then $a\sim t^{\de},\ \de>0$ (note that $a$ is analytic),
therefore, this implies $ \dot{a}\sim t^{\de-1} \mx{ and }
\ddot{a}\sim t^{\de-2}$, which in turn implies $R\sim t^{-2}$ and
hence it diverges.

Thus in a singularity free universe $a(t)$ is bounded from below, but
for condition I to hold we require $\ad\geq 0$ at all times. This
implies $a(t)\ra$ constant as $t\ra -\infty$, so that
$\ad,\ddot{a}\ra 0$ as $t\ra -\infty$. (See figure~\ref{figps2} for a possbile evolution of $a(t)$.)

The evolution of the scale factor reads, in the presence of a scalar field,
\begin{eqnarray}
\label{ST-Hubble}
H^2={\ad^2\over a^2}&=&\3(K+V)\equiv\3\rho_{\phi}\,, \\
\frac{\ddot{a}}{a}&=&\3(V-2K)\,,
\label{ST-accel}
\end{eqnarray}
where $K$ and $V$ are the kinetic and potential energies of the scalar
field. Since we argued that as $t\ra -\infty$, $\ad,\ddot{a}\ra 0$,
this then implies that $K,V\ra 0$ as $t\ra -\infty$, unless $a\ra 0$,
too. 

Consider the first case, when the ``particle'' starts out with
approximately zero kinetic and potential energies; then
\be
{d\rho_{\phi}\over dt}=-6H\dot{\phi}^2<0\,.
\ee
This leads to $\rho_{\phi}$ being negative at all finite times, but
this contradicts (\ref{ST-Hubble}).  We are left with the only
other possibility that $a\rightarrow 0$ as $t\ra -\infty$ such that
$\ad/a\ra H_0\sim {\rm constant}$. This means
\be
a(t)\st{t\ra -\infty}{\longrightarrow} a_0e^{H_0t}\,.
\ee
In other words we have shown that an exponential inflation is the only
way to realize the condition I. On the other hand it was shown in
\cite{Guth} that such spaces are geodesically incomplete, and this
is also the reason why it seems unlikely that such an evolution will
be able to resolve the Big Bang singularity. However, they may be
relevant for inflation. Therefore, let us proceed and try to pinpoint
the kind of $F(R)$ actions which can produce such an evolution.

The evolution equations for the scale factor (\ref{ST-Hubble})
and (\ref{ST-accel}) suggest
\be
K\st{t\ra -\infty}{\longrightarrow} 0\mx{ and }V\st{t\ra -\infty}
{\longrightarrow}\mt{constant}>0\,. 
\label{KV-assymp}
\ee
Specifically, since $\dot{\phi}\ra 0$, this means that $\ddot{\phi}\ra
0$, too. Then from the evolution equation for $\phi$:
\be
\ddot{\phi}+3H\dot{\phi}=-\frac{\p V}{\p \phi}
\Ra {\p V\over\p \phi}\st{t\ra -\infty}{\longrightarrow} 0\,.
\ee 
This is the key result which tells us that to obtain  an
exponentially expanding solution we need $\p V/\p \phi\ra 0$, a
profile which asymptotes to a flat potential. This may happen only if
$\partial V/\partial\phi\rightarrow 0$ as $\phi\rightarrow
\pm\infty$. This criterion will dictate what kind of $F(R)$ should
be chosen.

Let us first consider $F$ to be a finite series:
\be
F(R)=R+\sum_{n=0}^N c_n R^{n+2}\,,
\ee
with $N\geq 2$. Now, it is clear from (\ref{mapping}) that in
order for the mapping to work we need a sector of $F(R)$ where $F'(R)$
spans the interval $[0,\infty]$, so that $\phi$ spans
$(-\infty,\infty)$\footnote{In reality the $\phi$ field will be bounded
by some cut-off for the validity of the low energy effective theory to
work, but for the moment let us assume that $\phi$ can take
asymptotically large values.}.

\FIGURE[!h]{
\includegraphics[scale=0.6,angle=0]{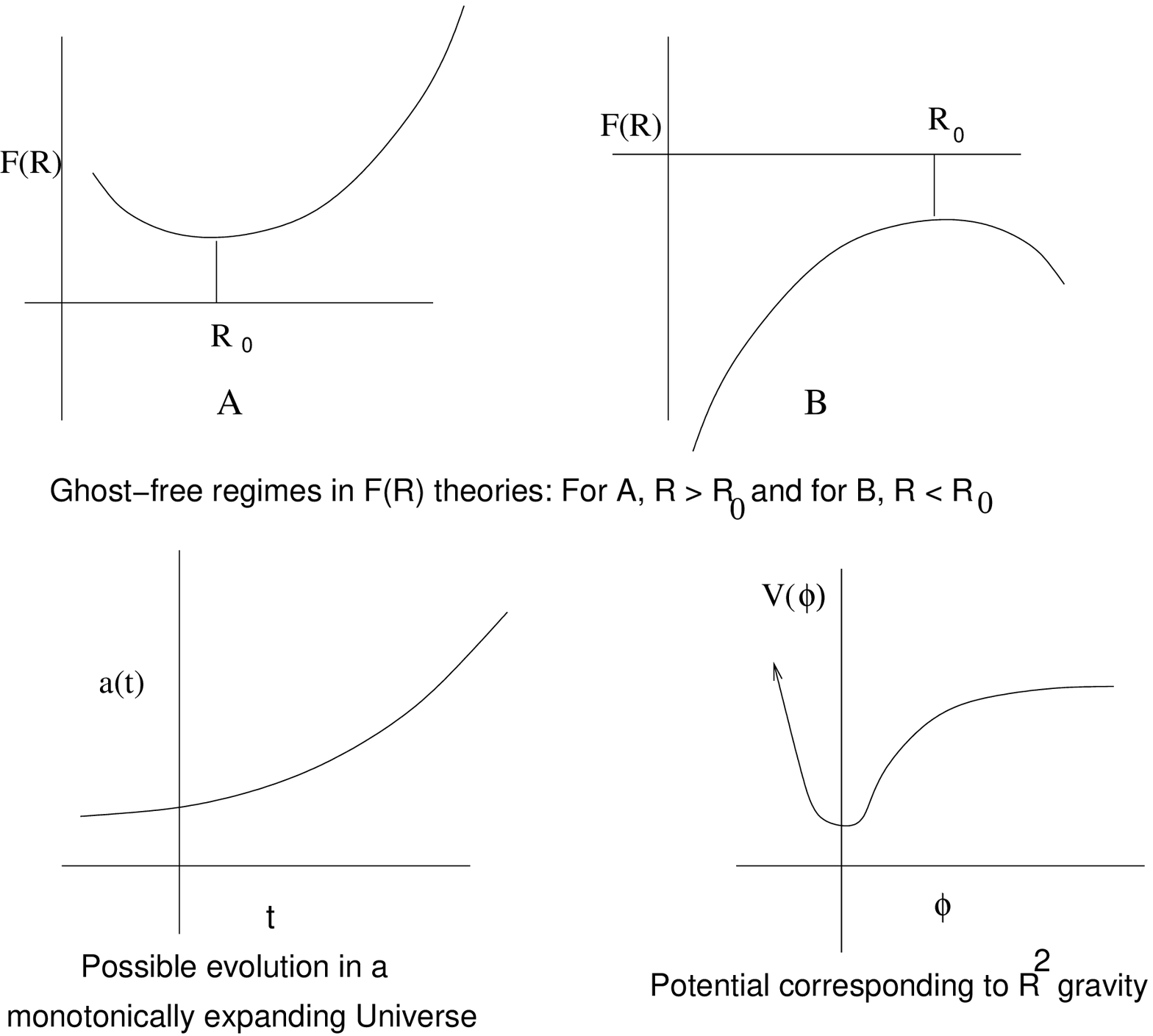}
%
\caption{\small Cosmology with $F(R)$ Theories}
\label{figps2}
}


There are two possible ways of realizing this criterion, which we have
qualitatively sketched  in Fig.~2. For concreteness let us
restrict our attention to Fig.~2A.  (Similar arguments can be made for
Fig.~2B.)  The relevant part of $F(R)$ that can realize this starts at
$R=R_0\neq 0$, which corresponds to $F'=0$ or $\phi\ra -\infty$, and
ends at $R\ra \infty$ corresponding to $F'\ra \infty$ or $\phi\ra
\infty$.

Note that we want to see whether $V'(\phi)\ra 0$ as
$\phi\ra\pm\infty$. Therefore we really need to look at the behavior
of $V(\phi)$ as $\phi\ra\pm\infty$. Now as $\phi\ra \infty$, $R\ra
\infty$, and therefore, $F(R)\ra c_NR^N$. The values of $R$ and $F(R)$
tend to:
\be
R\ra\left(\frac{1}{Nc_N}\right)^{\frac{1}{N-1}}
e^{\frac{\sqrt{2}}{\sqrt{3}(N-1)}\phi}~~\mx{ and }~~ 
F(R)\ra c_N\left(\frac{1}{Nc_N}\right)^{\frac{N}{N-1}}
e^{\frac{\sqrt{2}N}{\sqrt{3}(N-1)}\phi}\,.
\ee  
After some algebra we obtain (see also~\cite{zhuk}):
\be
V(\phi)\st{\phi\ra\infty}{\longrightarrow}\frac{N-1}{N}
\left(\frac{1}{Nc_N}\right)^{\frac{1}{N-1}}e^{-\al\phi}\,,~~~~
{\rm with}~~~~
\al=\frac{\sqrt{2}(N-2)}{\sqrt{3}(N-1)}<\sqrt{2}\,.
\label{exponent}
\ee
Except for $N=2$ this is an exponentially decaying function,
\ie, $V(\phi)\ra 0$ as $\phi\ra \infty$. Thus this cannot work, as we
required $V(\phi)$ to asymptote to a positive constant
(\ref{KV-assymp}).

Next, let us look at $\phi\ra -\infty$ or $R\ra R_0$, where
\be
F(R)\approx F(R_0)+\frac{F^I(R_0)}{I!}(R-R_0)^I\,,
\ee
where $F^I$ is the first non-zero derivative of $F$ at $R_0$. In this limit
\begin{eqnarray}
R\ra R_0+\left(\frac{I-1!}{F_I}\right)^{\frac{1}{I-1}}
e^{\frac{\sqrt{2}}{\sqrt{3}(I-1)}\phi}\,, \nonumber \\
F(R)\ra F(R_0)+\frac{F^I(R_0)}{I!}\left(\frac{I-1!}{F^I}\right)
^{\frac{I}{I-1}}e^{\frac{\sqrt{2}I}{\sqrt{3}(I-1)}\phi}\,.
\end{eqnarray}
In this case the asymptotic behavior of $\phi$ depends on whether
$F(R_0)$ is nonzero or not. Transforming $R,~F(R)$, we obtain
\be
V(\phi)\st{\phi\ra-\infty}{\longrightarrow}\left\{
\begin{array}{l}e^{-\frac{2\sqrt{2}}{\sqrt{3}}\phi} \mx{ if }
F(R_0)\neq 0\\ e^{-\frac{\sqrt{2}}{\sqrt{3}}\phi} \mx{ if } F(R_0)=0
\end{array}\right.\,.
\ee
As $\phi\ra-\infty$, the potential grows exponentially large
($V=\pm\infty$) rather than asymptoting to a constant as we want.
Therefore, a generic $F(R)$ cannot give rise to an exponential
expansion or inflation.

One exception arises (see (\ref{exponent})) when $F(R)\ra R^2$ as
$R\ra\infty$, \ie,
\be
F(R)=R+c_0R^2-2\La,.
\ee 
Indeed for such an action inflating solutions were found
earlier~\cite{Extended}.  Explicitly, one finds in this case
\be
V(\phi)=\frac{1}{4c_0}-\frac{1}{2c_0}
e^{-\frac{\sqrt{2}}{\sqrt{3}}\phi}+
\left(\frac{1}{4c_0}+2\La\right)e^{-\frac{2\sqrt{2}}{\sqrt{3}}\phi}\,.
\ee 
For $c_0>0$ and $1/4c_0+2\La>0$ this potential has the desired form to
give rise to inflation.

Even if one relaxes the condition that $F$ has to be a finite series,
the above conclusions essentially remain unchanged, as we will now
show. First note that the argument about the behavior of $V(\phi)$ as
$\phi\ra -\infty$ does not rely on the finiteness of the series,
but only on the analyticity of $F(R)$. Thus as $\phi\ra
-\infty$,~$V(\phi)$ exponentially rises as before and cannot satisfy
(\ref{KV-assymp}). We are left to consider what happens as
$\phi\ra \infty$. From the expression of the potential
(\ref{potential}), it is clear that if we want it to asymptote to
a constant then at least $R\ra e^{\frac{\sqrt{2}}{\sqrt{3}}\phi}$ or
$F(R)\ra e^{\frac{2\sqrt{2}}{\sqrt{3}}\phi}$ (or
faster). Consider the first case: This implies
\be
R\st{\phi\ra\infty}{\geq} \Phi=F'(R) \Ra F(R)\st{\phi\ra\infty}{\leq} R^2\,.
\ee
For the second case
\be
F(R)\st{\phi\ra\infty}{\geq} \Phi^2=F'^2(R)
\Ra F(R)\st{\phi\ra\infty}{\leq} R^2\,.
\ee
To have an exponentially inflating universe it is necessary
to have an $F(R)$ which asymptotes to $\pm R^2$ as
$R\ra\pm\infty$. Finally, we note in passing that adding
radiation/matter does not change this conclusion.


\section{Cosine Hyperbolic Bounces in $R+c_0R^2$ gravity}\label{Bouncepert}

Bounce solutions exist even when we consider only finite higher
derivative corrections. However, when at least for one $i>0$, $c_i\neq
0$ in (\ref{ordinary}), these theories necessarily have a ghost
and therefore we do not consider them any further.  Thus we are left
with the combination $F=R+c_0R^2$. In this case the trace equation is
given by (\ref{Gtrace}):
\be
\ti{G}=-\ti G_{00}+3\ti G_{ii}=-\left[6w-3w\left(1-6wc_0\right)
\mt{sech}^2\left(\sqw t\right)\right]\,=-4\La\,.
\ee
The bounce solution is then given by:
\be
{1\over w}=6c_0\,,~~~~ \mx{ and }~~~~ 4\La=6w={1\over c_0}\,.
\label{cosmo-relation}
\ee
We however expect from the previous arguments that such a bounce
solution is due to the ``implicit'' presence of ghost like
radiation. To see this, one computes $\ti{G}_{00}$. In particular, one finds
\be
L_{4}=-27 c_0w\mt{sech}^4\left(\sqw t\right)=
-{9\over 2}\mt{sech}^4\left(\sqw t\right)<0\,,
\ee
signaling a negative radiation energy density, or the presence of
ghosts.

We also note in passing that in the $R^2$ type gravity, to
get a bouncing solution, we require the presence of a typically large
(corresponding to the scale of higher curvature corrections)
cosmological constant. This is actually a generic feature of exact
hyperbolic cosine bounces, and we addressed this issue in
subsection~\ref{ETB}.

\section{Newtonian Limit}~\label{Newt}

To obtain the weak field limit, we look at the higher
curvature field equations directly. In \cite{schmidt} the field
equations were derived for a general $F(R,\Box R,\dots)$ type of
Lagrangian:
\begin{eqnarray}
\ti{G}_{\mu\nu}\equiv F_0R_{\mu\nu}-\2 Fg_{\mu\nu}-F_{0;\ \mu\nu}
+g_{\mu\nu}\Box F_0 \nonumber \\
+\2\sum_{m=1}^{\infty}\left[g_{\mu\nu}\left(F_m(\Box^{m-1}R)^
{;\ \s}\right)_{;\ \s}-F_{m;\ (\mu}(\Box^{m-1}R)_{;\ \nu)}\right]
=T_{\mu\nu}\,,
\label{generalized}
\end{eqnarray}
where 
\be
F_m\equiv \sum_{n=m}^{\infty}\Box^{n-m}{\p F\over \p \Box^n R}\,,
\ee
and as usual $T_{\mu\nu}$ is the matter stress-energy
tensor. It is easy to check that for ordinary Einstein gravity
$\ti{G}_{\mu\nu}$ reduces to the usual Einstein tensor. Now, in order
to derive the ``Newtonian potential'' corresponding to the scale
factor it is sufficient to just look at the trace equation, which
reads:
\be
\ti{G}\equiv g^{\mu\nu}\ti{G}_{\mu\nu}=F_0R-2F+3\Box F_0 
+2\sum_{m=1}^{\infty}F_m\Box^mR=-m\de(\vec{r})\,.
\label{G}
\ee 
where we have coupled gravity to a point particle source of mass $m$.
For $F$ of the form (\ref{quadratic}), one can show that:
\be
\ti{G}=G+\sum_m G_m\,,
\ee
where $G=g^{\mu\nu}G_{\mu\nu}=-R$, $G_{\mu\nu}$ being the usual
Einstein tensor, 
\begin{eqnarray}
G_0 &\equiv& 6c_0\Box R \,, \\
G_m &\equiv& c_m\left(6\Box^{i+1}R+2\sum_{n=1}^m\Box^{m-n}R\Box^nR\right)\
\forall\ m \neq 0\,. 
\end{eqnarray}
It is easy to check that $g_{\mu\nu}=\eta_{\mu\nu}$ is indeed the
right vacuum solution of the full higher curvature equations
(\ref{G}). Thus to obtain the Newtonian weak field limit we
have to expand around the flat space Minkowski metric,
\ie, $g_{\mu\nu}=\eta_{\mu\nu}+h_{\mu\nu}$; then
\be G=-\2\Box \hb\,, 
\ee
where 
\be 
\hb\equiv\hb^{\mu}{}_{\mu}\ ; \
\hb_{\mu\nu}=h_{\mu\nu}-\2\eta_{\mu\nu}h\,, 
\ee 
and we have used the Lorentz gauge $\hb_{\mu\nu}{}^{,\nu}=0$.

It is a straightforward task to derive the potential for the scale
factor,:  The expression for $\ti{G}$ simplifies to
\be 
\ti{G}=-\2 \Box\hb+3\sum_0^{\infty}c_i\Box^{i+2}\hb =
-\2\Box\Ga(\Box)\hb = -m\de(\vec{r})\,.
\ee 
Assuming $\hb$ to be slowly varying with time, we find it to be given
by:
$$
\hb(r)\sim \int d^3p\
{e^{i\vec{p}.\vec{r}}\over p^2\Ga(-p^2)} $$
$$
=\int_0^{\infty} dp \int_{-1}^1 d(\cos\te){p^2\over p^2\Ga(-p^2)}e^{ipr\cos\te}=\int_0^{\infty} dp {\over ipr\Ga(-p^2)}(e^{ipr}-e^{-ipr})$$
or 
\be
\hb(r)\sim {-i\over r}\int_{-\infty}^{\infty} dp {1\over p\Ga(-p^2)}e^{ipr}=
{-i\over r}\int_{-\infty}^{\infty} dp {e^{-\ga(-p^2)}\over p}e^{ipr}\,.
\ee
This completes our discussion on the Newtonian limit. 



\begin{thebibliography}{99}

\bibitem{WMAP}
D.~N.~Spergel {\it et al.}  [WMAP Collaboration],
  \newjournal{Astrophys.\ J.\ Suppl.}{APJSA}{148}{2003}{175},
  \astroph{0302209}.

\bibitem{Linde}
A. Linde, {\it Particle Physics and Inflationary Cosmology} (Harwood, Chur, 
Switzerland, 1990).

\bibitem{Brandenberger}
  V.~F.~Mukhanov, H.~A.~Feldman and R.~H.~Brandenberger,
  \prep{215}{1992}{203}.

\bibitem{Linde1}
  A.~D.~Linde, D.~A.~Linde and A.~Mezhlumian,
  \prd{49}{1994}{1783},
  \grqc{9306035}.

\bibitem{Borde}
A.~Borde and A.~Vilenkin,
  \grqc{9403004};
  \ijmpd{5}{1996}{813},
  \grqc{9612036}.

\bibitem{Guth}
 A.~Borde, A.~H.~Guth and A.~Vilenkin,
  \prl{90}{2003}{151301},
  \grqc{0110012}.

\bibitem{Hawking}
S. W. Hawking and G. F. R. Ellis, {\it The Large Scale Structure of Space-Time}
(Cambridge University, New York, 1975).

\bibitem{Starobinsky}
A. A. Starobinsky, \plb{91}{1980}{99}.

\bibitem{some}
 B.~Durin and B.~Pioline,
  \hepth{0501145};\\
  R.~Biswas, E.~Keski-Vakkuri, R.~G.~Leigh, S.~Nowling and E.~Sharpe,
  \jhep{0401}{2004}{064},
  \hepth{0304241};\\
 S.~Kawai, E.~Keski-Vakkuri, R.~G.~Leigh and S.~Nowling,
  \hepth{0507163};\\
  C.~V.~Johnson and H.~G.~Svendsen,
  \prd{70}{2004}{126011},
  \hepth{0405141};\\
  M.~Berkooz, B.~Pioline and M.~Rozali,
  \newjournal{JCAP}{JCAPA}{0408}{2004}{004},
  \hepth{0405126};\\
 C.~P.~Burgess, F.~Quevedo, S.~J.~Rey, G.~Tasinato and I.~Zavala,
  \jhep{0210}{2002}{028},
  \hepth{0207104};\\
  G.~T.~Horowitz and J.~Polchinski,
  \prd{66}{2002}{103512},
 \hepth{0206228};\\
  B.~Craps, D.~Kutasov and G.~Rajesh,
  \jhep{0206}{2002}{053},
  \hepth{0205101};\\
  S.~Elitzur, A.~Giveon, D.~Kutasov and E.~Rabinovici,
  \jhep{0206}{2002}{017},
  \hepth{0204189};\\
 F.~Larsen and F.~Wilczek,
  \prd{55}{1997}{4591},
  \hepth{9610252};\\
 A.~Lawrence,
  \jhep{0211}{2002}{019},
  \hepth{0205288};\\
  J.~Khoury, B.~A.~Ovrut, N.~Seiberg, P.~J.~Steinhardt and N.~Turok,
  \prd{65}{2002}{086007},
  \hepth{0108187};\\
  G.~Veneziano,
  \hepth{0002094};\\
G.~T.~Horowitz and A.~R.~Steif,
  \plb{258}{1991}{91};\\
  A.~J.~Tolley and N.~Turok,
  \prd{66}{2002}{106005},
  \hepth{0204091}.


\bibitem{Burgess}
 C.~P.~Burgess, F.~Quevedo, R.~Rabadan, G.~Tasinato and I.~Zavala,
  \newjournal{JCAP}{JCAPA}{0402}{2004}{008},
   \hepth{0310122}.

\bibitem{Stelle} 
K.~S.~Stelle,
  \prd{16}{1977}{953}.

\bibitem{Peter}
P.~Van Nieuwenhuizen,
  \npb{60}{1973}{478}.

\bibitem{Barton}
B.~Zwiebach, \plb{156}{1985}{315}.

\bibitem{hooft}
G. 't Hooft, \npb{72}{1974}{461}.

\bibitem{sft}
S. Mandelstam, \npb{64}{1973}{205};\\
M. Kaku and K. Kikkawa, \prd{10}{1974}{1110}, \prd{10}{1974}{1823};\\
M. Kaku, \prd{10}{1974}{3943};\\
E. Cremmer and J.-L. Gervais, \npb{76}{1974}{209}, \npb{90}{1975}{410};\\
M.B. Green and J.H. Schwarz, \npb{243}{1984}{475};\\
E. Witten, \npb{268}{1986}{253};\\
D.J. Gross and A. Jevicki, \npb{283}{1987}{1}, \npb{287}{1987}{225};\\
Z. Hlousek and A. Jevicki, \npb{288}{1987}{131};\\
E. Cremmer, C.B. Thorn, and A. Schwimmer, \plb{179}{1986}{57};\\
C.B. Thorn, The oscillator representation of Witten's three open string vertex function, in {\it Proc. XXIII Int. Conf. on High Energy Physics}, July 16-23, 1986, Berkeley, 
CA, ed. S.C. Loken (World Scientific, Singapore, 1987) v.1, p. 374. 

\bibitem{many}
  P.~J.~Steinhardt and N.~Turok,
  \hepth{0111030};\\
  F.~Di Marco, F.~Finelli and R.~Brandenberger,
  \prd{67}{2003}{063512},
  \astroph{0211276};\\
  J.~E.~Lidsey, D.~J.~Mulryne, N.~J.~Nunes and R.~Tavakol,
  \prd{70}{2004}{063521},
  \grqc{0406042};\\
R.~H.~Brandenberger, V.~Mukhanov and A.~Sornborger,
  \prd{48}{1993}{1629},
  \grqc{9303001};\\
 R.~H.~Brandenberger, R.~Easther and J.~Maia,
 \jhep{9808}{1998}{007},
  \grqc{9806111};\\
D.~A.~Easson and R.~H.~Brandenberger,
  \jhep{9909}{1999}{003},
  \hepth{9905175};\\
 M.~Gasperini, M.~Maggiore and G.~Veneziano,
  \npb{494}{1997}{315},
  \hepth{9611039};\\
  R.~Easther and K.~i.~Maeda,
  \prd{54}{1996}{7252},
  \hepth{9605173}.

\bibitem{jim} 
J.~M.~Cline, S.~Y.~Jeon and G.~D.~Moore,
  \prd{70}{2004}{043543},
  \hepph{0311312}.

\bibitem{Weinberg} 
S. Weinberg, Critical phenomena for field theorists, in {\it Understanding the fundamental 
constituents of matter}, proc. of the International School of Subnuclear Physics, Erice, 
1976, ed. A. Zichichi (Plenum, New York, 1978) p. 1;\\
Ultraviolet divergences in quantum theories of gravitation, in {\it General relativity: an 
Einstein centenary survey}, eds. S.W. Hawking and W. Israel (Cambridge University, 
Cambridge, 1979) p. 790.

\bibitem{warren} 
 W.~Siegel,
  \hepth{0309093}.

\bibitem{solganik}
 A.~Nunez and S.~Solganik,
  \plb{608}{2005}{189},
  \hepth{0411102};\\
  \hepth{0403159};\\
T.~Chiba,
  \newjournal{JCAP}{JCAPA}{0503}{2005}{008},
\grqc{0502070}.

\bibitem{Wands}
D.~Wands,
  \cqg{11}{1994}{269},
  \grqc{9307034}.

\bibitem{Will}
 C.~M.~Will,
  \newjournal{Living Rev.\ Rel.}{00222}{4}{2001}{4},
  \grqc{0103036}.

\bibitem{Andrew}
 A.~R.~Liddle, A.~Mazumdar and J.~D.~Barrow,
  \prd{58}{1998}{027302},
  \astroph{9802133}.

\bibitem{Extended}
F.~S.~Accetta and J.~J.~Trester,
  \prd{39}{1989}{2854};\\
E.~J.~Weinberg,
  \prd{40}{1989}{3950};\\
D.~La, P.~J.~Steinhardt and E.~W.~Bertschinger,
  \plb{231}{1989}{231};\\
R.~Holman, E.~W.~Kolb and Y.~Wang,
  \prl{65}{1990}{17};\\
A.~D.~Linde,
  \plb{238}{1990}{160};\\
B.~A.~Campbell, A.~D.~Linde and K.~A.~Olive,
  \npb{355}{1991}{146};\\
J.~D.~Barrow and P.~Parsons,
  \prd{55}{1997}{1906},
  \grqc{9607072};\\
M.~Susperregi and A.~Mazumdar,
  \prd{58}{1998}{083512},
  \grqc{9804081};\\
J.~Garcia-Bellido, A.~D.~Linde and D.~A.~Linde,
  \prd{50}{1994}{730},
  \astroph{9312039};\\
A.~Mazumdar,
  \plb{469}{1999}{55},
  \hepph{9902381};\\
A.~Mazumdar and J.~Wang,
  \plb{490}{2000}{251},
  \grqc{0004030};\\
L.~E.~Mendes and A.~Mazumdar,
  \plb{501}{2001}{249},
 \grqc{0009017};\\
A.~A.~Starobinsky, S.~Tsujikawa and J.~Yokoyama,
  \npb {610} {2001}{383}
  \astroph{0107555};\\
A.~A.~Starobinsky and J.~Yokoyama,
  \grqc{9502002};\\
 A.~M.~Green and A.~R.~Liddle,
  \prd {54}{1996}{2557},
  \astroph{9604001};\\
A.~Mazumdar and A.~Perez-Lorenzana,
  \plb{508}{2001}{340},
  \hepph{0102174};\\
A.~Mazumdar, R.~N.~Mohapatra and A.~Perez-Lorenzana,
 \newjournal{JCAP}{JCAPA}{0406}{2004}{004}
  \hepph{0310258};\\
T.~Biswas, R.~Brandenberger, D.~A.~Easson and A.~Mazumdar,
  \prd{71}{2005}{083514},
  \hepth{0501194}.

\bibitem{Copeland}
 P.~M.~Saffin, A.~Mazumdar and E.~J.~Copeland,
  \plb{435}{1998}{19},
  \hepth{9805032}.

\bibitem{Mendes}
A.~Mazumdar and L.~E.~Mendes,
  \prd{60}{1999}{103513},
  \hepph{9902274}.

\bibitem{Us}
  P.~J.~E.~Peebles and B.~Ratra,
  \apj{325}{1988}{L17};\\
 D.~F.~Torres,
  \prd{66}{2002}{043522},
  \astroph{0204504};\\
T.~Biswas and A.~Mazumdar,
  \hepth{0408026};\\
T.~Biswas, R.~Brandenberger, A.~Mazumdar and T.~Multamaki,
  \hepth{0507199};\\
S.~M.~Carroll, A.~De Felice, V.~Duvvuri, D.~A.~Easson, M.~Trodden and M.~S.~Turner,
  \prd{71}{2005}{063513},
  \astroph{0410031}.

\bibitem{magnano} 
 G.~Magnano and L.~M.~Sokolowski,
  \prd{50}{1994}{5039},
 \grqc{9312008}.

\bibitem{marc} 
T.~Biswas, M.~Grisaru and W.~Siegel,
  \npb{708}{2005}{317},
  \hepth{040908}.

\bibitem{expo}
  D.~Evens, J.~W.~Moffat, G.~Kleppe and R.~P.~Woodard,
  \prd{43}{1991}{499};\\
A.A. Tseytlin, 
\plb{363}{1995}{223},
\hepth{9509050}.

\bibitem{mcinnes}
  B.~McInnes,
  \jhep{0208}{2002}{029},
  \hepth{0112066}.

\bibitem{zhuk}
  U.~Gunther, A.~Zhuk, V.~B.~Bezerra and C.~Romero,
  \cqg{22}{2005}{3135},
  \hepth{0409112}.
  
\bibitem{schmidt} 
H.~J.~Schmidt,
  \cqg{7}{1990}{1023}.

\end{thebibliography}
\end{document}